\documentclass[epj]{svjour}
\usepackage{amssymb}
\usepackage{amsmath}
\usepackage{graphicx}

\begin{document}
\title{Deterministic instabilities in the magneto-optical trap}
\author{Andrea di Stefano, Philippe Verkerk and Daniel Hennequin}
\institute{Laboratoire de Physique des Lasers, Atomes et Mol\'{e}cules,\\
Unit\'{e} mixte du Centre National de la Recherche Scientifique,\\
Centre d'Etudes et de Recherches Lasers et Applications, \\
B\^{a}t. P5, Universit\'{e} des Sciences et Technologies de Lille,\\
F-59655 Villeneuve d'Ascq cedex - France}
\date{\today }

\abstract{
The cloud of cold atoms obtained from a magneto-optical trap is known to
exhibit two types of instabilities in the regime of high atomic densities:
stochastic instabilities and deterministic instabilities. In the present
paper, the experimentally observed deterministic dynamics is described
extensively. Three different behaviors are distinguished. All are cyclic,
but not necessarily periodic. Indeed, some instabilities exhibit a cyclic
behavior with an erratic return time. A one-dimensional stochastic model
taking into account the shadow effect is shown to be able to reproduce the
experimental behavior, linking the instabilities to a several bifurcations.
Erraticity of some of the regimes is shown to be induced by noise.
}
\maketitle

\section{Introduction}

Experimental quantum physics knows since several years spectacular results,
thanks to a simplification to produce quantum objects with long coherence
times or macroscopic dimensions. Let us cite the achievement of the
Bose-Einstein condensates\cite{nobel}, the characterization of quantum chaos 
\cite{QC}, the improvement of atomic clocks\cite{clocks}, the designs of
quantum computers and quantum communication systems\cite{qcomp}, or also the
accurate understanding of quantum decoherence\cite{deco}. One of the basic
tools used to obtain most of these results is the Magneto-Optical Trap
(MOT), which performs the cooling of atoms at temperatures of the order of
the $\mu $K: this is the first step before reaching lower temperatures where
the quantum properties of atoms dominate. Although it is a key device in the
new atomic physics, the basic mechanisms determining the properties of the
cloud of cold atoms in a MOT have been poorly studied, and the collective
dynamics of these still ``classical'' atoms have been almost ignored, even
though the existence of instabilities in the MOT is known since the first
realizations. On the contrary, some simple empirical rules are used to avoid
these inconveniences. Nevertheless, an accurate knowledge of the individual
and collective behaviors of the cold atoms in the cloud could help in
understanding the limitations of the process, and above all, to enhance it
through the control of the dynamics, as it was done in many other systems,
in physics and other fields of science.

However, a necessary preamble to such applications is the identification of
the nature of the dynamics observed in MOTs. Indeed, complex behaviors may
be subdivided in two groups: stochastic and deterministic behaviors. For the
former, the dynamics originate in noise, i.e. in dynamical components,
usually with a large number of degrees of freedom, considered as external to
the system. This is usually experimental technical noise, and requires to
add in the model a random component. Such a complex dynamics is meaningless
from the physical point of view, because it cannot give any new informations
about the MOT mechanisms. On the contrary, deterministic dynamics are
intrinsic to the system, and do not require to add anything to the model:
periodic instabilities can appear with two degrees of freedom, while chaos
needs at least three degrees of freedom. This last case opens many
perspectives: for example, it is possible to reach new working points by the
methods of control of chaos, or to measure parameters which are usually
inaccessible\cite{bistab}.

Recent studies have shown that the collective behavior of the atomic cloud
produced by a MOT exhibit both stochastic instabilities\cite%
{nousprl,bruit2002} and deterministic instabilities\cite{InstDet}. The
former has been extensively described in \cite{bruit2002}. A model
demonstrates that the different stochastic behaviors observed in the
experiments are well explained if the absorption of light by the atoms is
taken into account, through the so-called shadow effect \cite{shadow}. It is
also shown that these stochastic instabilities are not ``instabilities'' in
the usual meaning, as they result from an amplification of noise, due, from
a dynamical point of view, to the folded structure of the stationary
solutions. The same model was also predicting, for slighly different values
of the parameters, deterministic instabilities which, in turn, has been
observed experimentally\cite{InstDet}.

In the present paper, we report an extensive study of these deterministic
instabilities. We detail and complete the experimental results given in \cite%
{InstDet}, and analyze accurately the mechanisms leading to the different
deterministic regimes through the model introduced in \cite{InstDet}. We
show in particular that the model is able to reproduce each type of
dynamics, and predicts deterministic chaos. We show also the main role that
noise is still playing in the dynamics.

The paper is organized as follows. After this introduction, section \ref%
{S2ExpSetup} describes briefly the experimental setup, and section \ref%
{S3ExpRes} gives a detailed analysis of the experimental observations.
Section \ref{S4DetMod} is devoted to a short description of the model,
already detailed in \cite{bruit2002}. In section \ref{S5StatSol}, the
stationary solutions of the model are discussed, and in section \ref{S6Dyn},
the deterministic dynamical behavior predicted by the model is described,
and compared with the experiment results. Finally, in section \ref{S7Noise},
the effect of noise on the dynamics is studied.

\section{Experimental set-up}

\label{S2ExpSetup}The experimental set-up has already been described in
detail elsewhere\cite{bruit2002}, and thus the description here is
simplified. The Cesium-atom MOT is in the usual $\sigma _{+}-\sigma _{-}$\
configuration, with three arms of two counter-propagating beams obtained
from the same laser diode. The waist $w_{T}$ of the trap beams may be varied
from typically $3$ to $10$~mm. We use a configuration where
counter-propagating beams result from the reflection of the three forward
beams. This simplifies the detection of the dynamics, as compared to a six
independent beams configuration. Indeed, because of the shadow effect, a
center-of-mass motion is generated. However, as the nonlinearities involved
in both cases are the same, we expect that the dynamics will be
fundamentally of the same nature in the two configurations.

\begin{figure}[tph]
\centerline{\resizebox{0.45\textwidth}{!}{\includegraphics{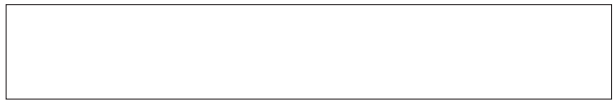}}}
\caption{Sequence of snapshots showing the time evolution of the unstable
atomic cloud. Snapshots are presented in the chronological order, each one
being separated by 40 ms. This sequence corresponds to the fast stage of a C$%
_{P}$ cycle (see section \protect\ref{S3ExpRes}).}
\label{fig:film}
\end{figure}

The dynamics of the atomic cloud consists in a deformation of the spatial
atomic distribution, leading to fluctuations of the shape of the cloud, as
illustrated in Fig. \ref{fig:film}. Therefore, the relevant {\em dynamical
variables} allowing us to describe instabilities, could be the shape of the
cloud (i.e. for example the {\em local} velocities and atomic densities in
the cloud). This type of description corresponds to a high dimensional
model, associated with partial differential equations. Here, for the sake of
simplicity, we choose to limit our description to the center of mass (CM)
location ${\bf r}$, and the {\em total} number of atoms $n$ in the atomic
cloud. This allows us to model the system with ordinary differential
equations, and reduces the dimension to seven, and even three in a 1D model.
As it is shown in the following, the use of this description appears to be
sufficient to understand the main mechanisms of the instabilities.

A 4-quadrant photodiode (4QP) is used to detect the fluorescence of the
cloud. The differential signal of the 4QP allows us to measure the motion of
the CM through one of its component $r$, while the total signal gives us the
number $n$ of atoms inside the cloud. A second 4QP, perpendicular to the
first one, prevents the measure from line-of-sight effects due to the
optical thickness of the cloud. We checked that whatever the type of
dynamical behavior, the signal recorded by both 4QP have the same properties
and are qualitatively identical.

Parameters acting on the dynamics have been extensively discussed in \cite%
{bruit2002}. The detuning $\Delta _{0}$ of the MOT, the magnetic field
gradient $G$, the MOT beam intensity $I_{1}$ and the repumper laser
intensity $I_{rep}$ may be considered as {\em control parameters}, because
they can be easily changed in the experiment. On the contrary, the alignment
of the MOT beams, the vapor pressure in the cell and the MOT beam waist,
which also play a crucial role in the dynamics, cannot be considered as
control parameters, either because they cannot be changed easily, or because
they cannot be measured with accuracy. Therefore, these parameters have not
been varied in the experiments. The parameter ranges explored in the present
experiment are summarized in Tab. \ref{tabexpparams}.

\begin{table}[h]
\caption{Range of the parameters used in the present experiment. $G$ is the
magnetic field gradient, $I_{+}$ is the intensity of the forward beam and $%
\protect\delta $ is the detuning. $I_{s}$ is the saturation intensity ( $%
I_{s}=1.1$ mW) and $\Gamma $ is the natural width of the transition. The
last column indicates the default parameter values used to obtain the
results reported in the present paper.}
\label{tabexpparams}%
\begin{tabular}{rcc}
& range & default set \\ \hline
$G$ & $G\leq 14$ Gcm$^{-1}$ & 14 Gcm$^{-1}$ \\ 
$I_{+}=I/I_{s}$ & $4\leq I_{+}\leq 20$ & 10 \\ 
$\Delta =\delta /\Gamma $ & $\Delta \leq -0.5$ & -%
\end{tabular}%
\end{table}

\section{Experimental results}

\label{S3ExpRes}In \cite{bruit2002}, it has been shown that the atomic cloud
exhibits two types of instabilities, depending on the parameters of the MOT,
in particular the trap laser beam intensity $I_{1}$. When $I_{1}$ is small,
typically less than $10I_{S}$ ($I_{S}=1.1$~mW/cm$^{2}$ is the saturation
intensity), instabilities are essentially stochastic. Depending on the other
parameters, as the detuning, several types of stochastic instabilities
occur. In the {\it S}$_{{\it L}}$ behavior, instabilities are characterized
by a unique slow time scale, and no component larger than 2 Hz appears
neither in the motion of the trap, nor in its population. On the contrary,
in the {\it S}$_{{\it H}}$ behavior, a second time scale, at higher
frequency (typically from 20 to 100 Hz) appear in the trap motion, but not
in the population\cite{bruit2002}.

When $I_{1}$ is increased, {\it S} instabilities are progressively replaced
by {\it C} instabilities ({\it C} stands for cyclic). These instabilities
have already been described in \cite{InstDet} for a given set of parameters.
In the following, we extend this description for the whole range of
parameters where such instabilities appear. We also discuss in more details
than in \cite{bruit2002}, the connections between {\it S} and {\it C}\
instabilities, in particular through their respective domain of appearance.

All {\it C} behaviors that we observed in the experiments have in common to
have a large amplitude, and to be cyclic, i.e. their trajectory in the phase
space follows a close cycle. In the time domain, the signal exhibits the
same pattern, which is repeated indefinitely. However, the cadence of the
signal is not necessarily periodic, but can be erratic. Thus different types
of {\it C} instabilities may be distinguished, and we arbitrarily classified
them into three groups, that we call {\it C}$_{{\it P}}$, {\it C}$_{{\it 1}}$
and {\it C}$_{{\it S}}$ instabilities.

\begin{figure}[tph]
\centerline{\resizebox{0.45\textwidth}{!}{\includegraphics{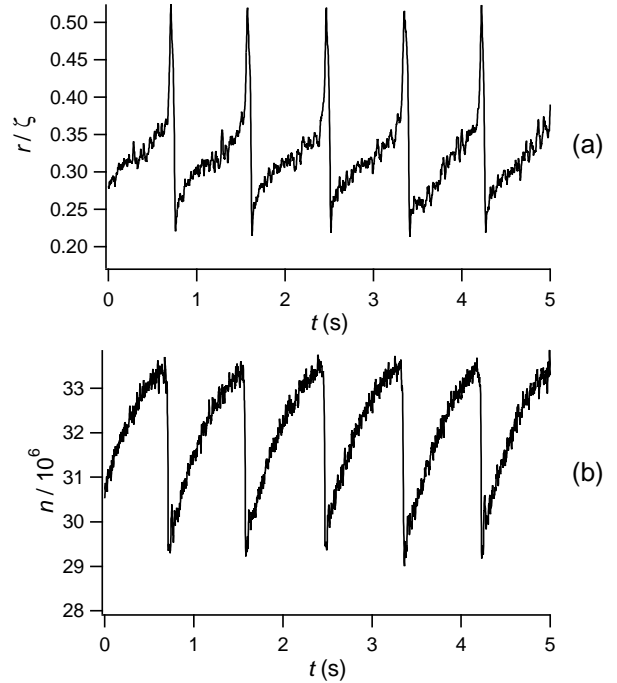}}}
\caption{Experimental record of the time evolution of the cloud when {\it {C$%
_P$}} instabilities occur. In (a), CM motion $r$; in (b) population $n$. $%
\Delta_0=-1.40$, $I_1=10$ and $I_rep=1.2$ mW/cm$^2$.}
\label{fig:cp}
\end{figure}

Among all types of instabilities observed in the MOT, {\it C}$_{{\it P}}$
instabilities are the most typical deterministic behavior (Fig. \ref{fig:cp}%
). Indeed, they are characterized by periodic oscillations, with a frequency
of the order of 1 Hz and a large motion amplitude of the order of 100 $\mu $%
m to 1 mm, while the population variations are typically 10\thinspace \%.
The main feature of the cycle is its asymmetry, which can be described by
the succession of two stages with different durations. During the long
stage, $r$ and $n$ behave in the same way, increasing slowly on a
significant amplitude, which represents about 30\% of the full $r$
amplitude, and 100\% of the full $n$ amplitude. During the short stage, $r$
and $n$ change rapidly: $n$ decreases to come back to the initial value of
the long stage, while $r$ makes a fast oscillation, with an amplitude much
larger than during the long stage. This means that the two stages are not
only different by their duration, but also by the dynamical time scales,
much faster during the short stage. In fact, the characteristic time of the
dynamics during the short stage is more than one order of magnitude smaller
than that in the long stage.

\begin{figure}[tph]
\centerline{\resizebox{0.45\textwidth}{!}{\includegraphics{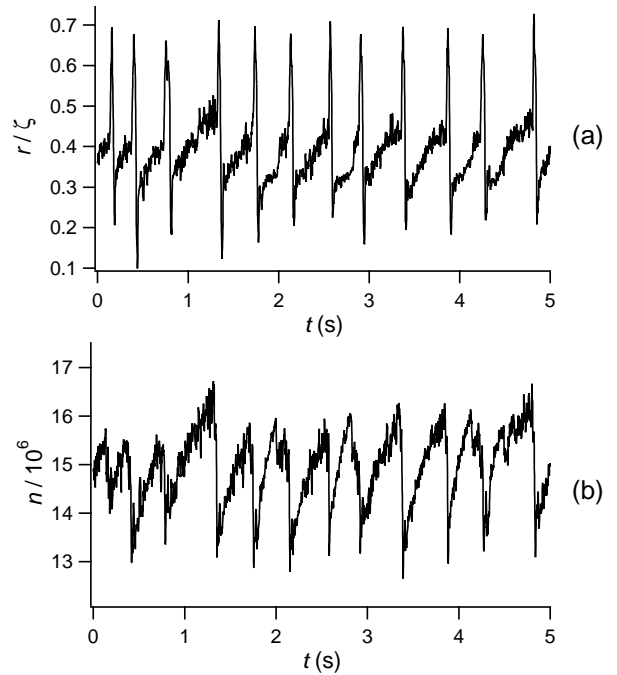}}}
\caption{Experimental record of the time evolution of the cloud when {\it {C$%
_1$}} instabilities occur. In (a), CM motion $r$; in (b) population $n$. $%
\Delta_0=-0.85$, $I_1=10$ and $I_rep=1.2$ mW/cm$^2$.}
\label{fig:c1}
\end{figure}

{\it C}$_{{\it 1}}$ instabilities, illustrated in Fig. \ref{fig:c1},
corresponds to a motion of the cloud very similar to that of {\it C}$_{{\it P%
}}$ instabilities. Indeed, $r$ exhibits the same behavior along the same
type of cycle, covered with two different time scales separated by one order
of magnitude. The main difference comes from the erraticity of the motion,
which is no more periodic: indeed, although the basic pattern of the motion
remains the same cycle, the duration of each cycle fluctuates. From the
dynamical point of view, it is convenient use the {\em return time} between
two cycles, which, in chaotic dynamics, is known to be a relevant variable
of the system\cite{RT}. The $r$ return time is constant in periodic motions (%
{\it C}$_{{\it P}}$ instabilities), while it is fluctuating in {\it C}$_{%
{\it 1}}$ instabilities. The $n$ return time is also varying in {\it C}$_{%
{\it 1}}$ instabilities, but other differences as compared to {\it C}$_{{\it %
P}}$ regime appear. In particular, the ratio between the fast and slow
stages of the dynamics fluctuates, so that for some periods, the two stages
occur with a comparable time scale. In short, {\it C}$_{{\it 1}}$
instabilities appears as {\it C}$_{{\it P}}$ instabilities, in which noisy
fluctuations appear, essentially on the return time.

\begin{figure}[tph]
\centerline{\resizebox{0.45\textwidth}{!}{\includegraphics{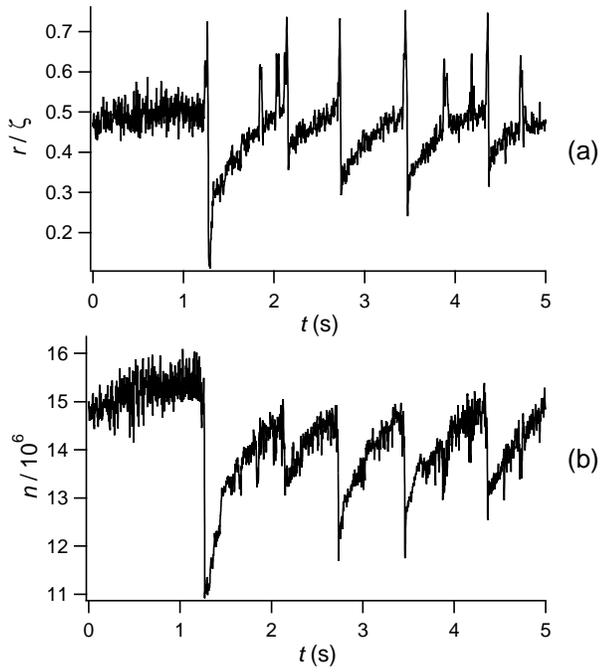}}}
\caption{Experimental record of the time evolution of the cloud when {\it {C$%
_S$}} instabilities occur. In (a), CM motion $r$; in (b) population $n$. $%
\Delta_0=-0.75$, $I_1=10$ and $I_rep=1.2$ mW/cm$^2$.}
\label{fig:cs}
\end{figure}

With {\it C}$_{{\it S}}$ instabilities (Fig. \ref{fig:cs}), any periodicity
has disappeared from the behavior of both $r$ and $n$. Not only the return
time is fluctuating, but also the amplitude of the cycles changes with time.
In fact, speaking of cyclic behavior in the present case is excessive,
except that the basic pattern of the motion keeps similarities with the
cycle of {\it C}$_{{\it P}}$ instabilities, in particular the two stages
with different time scales. However, even the motion cycle is irregular,
with secondary fast oscillations appearing during the slow stage, while the
amplitude of the $n$ cycle can fluctuate in a ratio of 1:5.

\begin{figure}[tph]
\centerline{\resizebox{0.45\textwidth}{!}{\includegraphics{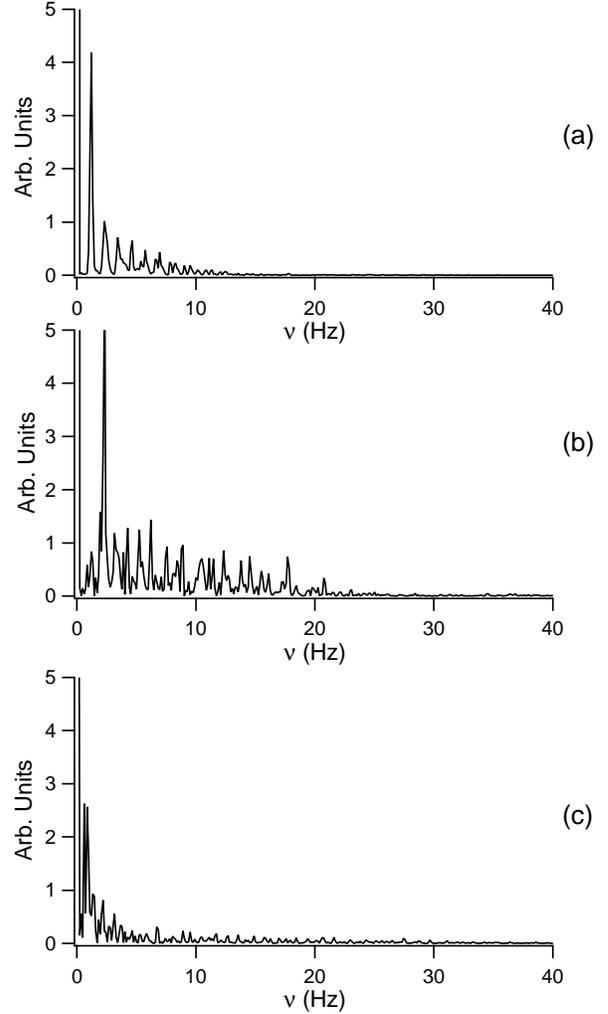}}}
\caption{Spectra of the CM location time evolution in case of {\it {C}}
instabilities. Figures (a), (b) and (c) correspond respectively to Fig. 
\protect\ref{fig:cp}a, Fig. \protect\ref{fig:c1}a and Fig. \protect\ref%
{fig:cs}a.}
\label{fig:VII5}
\end{figure}

The differences between the three {\it C} behaviors are particularly well
illustrated by the power spectrum of $r$, as shown in Fig. \ref{fig:VII5}.
The spectrum of {\it C}$_{{\it P}}$ instabilities exhibits a first large and
narrow peak, at about 1 Hz, corresponding to the main period of the signal,
followed by a series of harmonics (Fig. \ref{fig:VII5}a). These harmonics
are a signature of the second time scale, much faster than the basic period,
which appears in the fast oscillation of $r$. For {\it C}$_{{\it 1}}$
instabilities (Fig. \ref{fig:VII5}b), the first peak remains, demonstrating
that the signal remains essentially periodic. However, the regularly spaced
harmonics have disappeared, but high frequency components remain. They are
distributed erratically, but have globally a larger weight than in {\it C}$_{%
{\it P}}$ instabilities. Finally, for {\it C}$_{{\it S}}$ instabilities, the
main frequency component has decreased drastically, and the spectrum may
rather be considered as a wide spectrum, as those observed in chaotic or
stochastic signals. Unfortunately, it is impossible to distinguish between
these two possibilities through the spectrum analysis of the behavior. To do
so, it is necessary to turn to more powerful techniques, such as the
reconstruction of the attractor of the dynamics.

\begin{figure}[tph]
\centerline{\resizebox{0.45\textwidth}{!}{\includegraphics{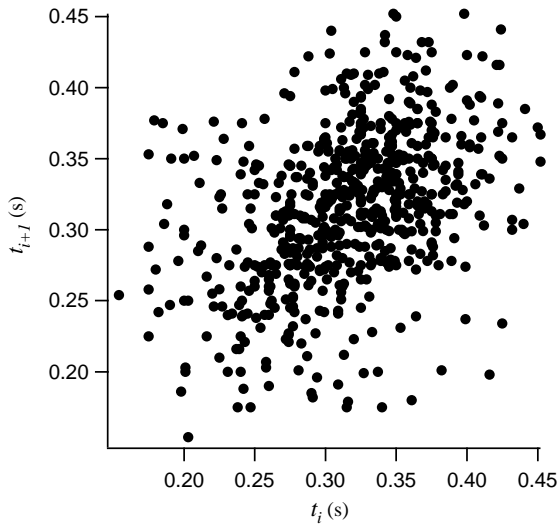}}}
\caption{Typical first return time diagram, obtained from a time series of
200 s with $\Delta _{0}=-1$, $I_{1}=16$ and $I_{rep}=7.5$ mW/cm$^{2}$.}
\label{fig:return}
\end{figure}

Indeed, the MOT is a dissipative system, and any deterministic behavior lies
in the phase space on an attractor. In the case of deterministic chaos, this
attractor is complex, but has usually a structured shape, easily
recognizable. On the contrary, if the behavior is dominated by noise, there
is no attractor, and the trajectories fill the whole phase space. The
reconstruction of the attractor of a system from experimental time series is
a well mastered operation. It can be performed following several methods
(delays, derivatives), and needs usually additional steps, as the plot of
the Poincar\'{e} section. In the present case, the use of return times is
particularly well adapted, as it appears as one of the main properties of
the behavior. The return time diagram, which is equivalent to a Poincar\'{e}
section, consists in plotting the return time between cycles $n$ and $n+1$,
as a function of the return time between the cycles $n-1$ and $n$\cite{RTD}.
The Poincar\'{e} section is a cross-section of the attractor, and has a
dimension decreased of one unit as compared to the attractor. Thus, first
return time diagrams of {\it C}$_{{\it P}}$ instabilities give just a point,
as expected from a cyclic behavior. Fig. \ref{fig:return} shows the first
return time diagram in the case of {\it C}$_{{\it 1}}$ instabilities. It is
a good illustration of all the return times diagram we have obtained for 
{\it C}$_{{\it 1}}$ and {\it C}$_{{\it S}}$ instabilities: points appear to
be distributed randomly in the phase space, without any structure. Thus we
can conclude that the behavior is either high-dimensional deterministic
chaos, or a stochastic dynamics. From the point of view of the model
discussed below, these two hypotheses are equivalent, as the model includes
only three degrees of freedom.

\begin{figure}[tph]
\centerline{\resizebox{0.45\textwidth}{!}{\includegraphics{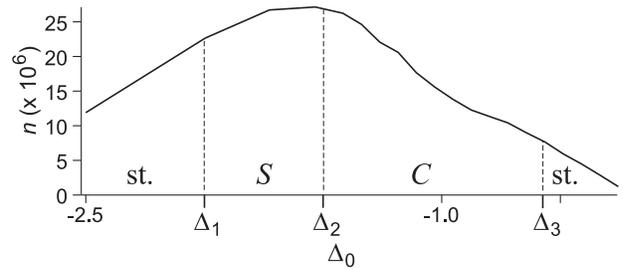}}}
\caption{This figure illustrates the evolution of the behavior as a function
of the detuning for $I_{1}=6.8$ and $I_{rep}=1.5$ mW/cm$^{2}$; The full line
reports the population, while the dashed lines separate the domain of
different behaviors: st. stands for stable, S (C) for $S$ ($C$)
instabilities.}
\label{fig:VII3}
\end{figure}

\begin{figure}[tph]
\centerline{\resizebox{0.45\textwidth}{!}{\includegraphics{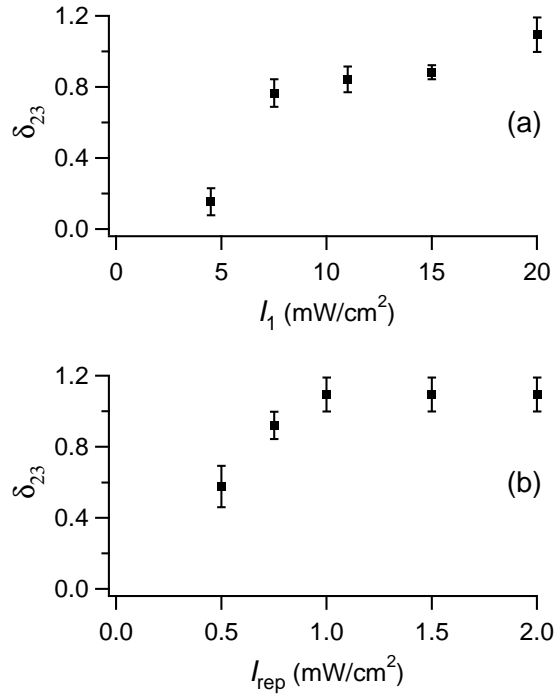}}}
\caption{Width $\protect\delta _{23}$ of the {\it {C} instabilities area as
a function of (a) the MOT intensity $I_{1}$ and (b) the repumper intensity $%
I_{rep}$. In (a), $I_{rep}=1.5$ mW/cm$^{2}$, and in (b), $I_{1}=18$}}
\label{fig:largeurvsI}
\end{figure}

The links between the different {\it C} regimes appear clearly when one
looks at the dependence of the behavior versus the different control
parameters, and in particular $I_{1}$ and $\Delta _{0}$. It was already
shown in \cite{bruit2002} that for small trap beam intensity $I_{1}$
(typically $I_{1}\leq 3$), the atomic cloud exhibits only {\it S}
instabilities. When $I_{1}$ is increased, {\it S} instabilities still exist,
but they are progressively superseded by {\it C} instabilities. The
appearance of {\it C} instabilities occurs progressively, at the cost of 
{\it S} instabilities. For intermediate values of $I_{1}$, both types of
instabilities exist. Their typical distribution versus $\Delta _{0}$ is
illustrated in Fig. \ref{fig:VII3}: far from resonance, the cloud is stable;
as the resonance is approached, {\it S} instabilities appear for a detuning $%
\Delta _{0}=\Delta _{1}$. Then {\it C} instabilities appear in $\Delta
_{2}>\Delta _{1}$. If the detuning is still increased, {\it C} instabilities
disappear in $\Delta _{3}$ at the benefit of a stable behavior. Finally, the
cloud vanishes in $\Delta _{4}$. As $I_{1}$ is increased, the interval $%
\delta _{23}=\Delta _{3}-\Delta _{2}$ where {\it C} instabilities occur,
increases at the cost of the width $\delta _{12}=\Delta _{2}-\Delta _{1}$
where {\it S} instabilities occur, while the total unstable interval $\delta
_{13}=\Delta _{3}-\Delta _{1}$ remains more or less constant. When {\it C}
instabilities merge for $I_{1}=4I_{S}$, they appear on\ a narrow interval $%
\delta _{23}\gtrsim 0$ (Fig. \ref{fig:largeurvsI}). This interval increases
rapidly until $I_{1}=7.5I_{S}$ and $\delta _{23}=0.8$. For $I_{1}>7.5I_{S}$, 
$\delta _{23}$ increases more slowly, to reach the value of $\delta _{23}=1$
in $I_{1}=20I_{S}$. The value of $\delta _{23}$ depends also on the other
parameters of the system. Fig. \ref{fig:largeurvsI}b illustrates, as an
example, how it depends on the repumper intensity $I_{rep}$, at given $I_{1}$%
: $\delta _{23}$ varies from 0 for $I_{rep}\simeq 0.4$~mW/cm$^{2}$ to $1$
for $I_{rep}>1$~mW/cm$^{2}$.

Between $\Delta _{2}$ and $\Delta _{3}$, the different types of {\it C}
instabilities appear, following a constant scenario. In $\Delta _{2}$, when
they merge abruply from a stable or {\it S} behavior, they are {\it C}$_{%
{\it P}}$ instabilities, and as $\Delta _{0}$ is increased, they transform
successively in {\it C}$_{{\it 1}}$, then {\it C}$_{{\it S}}$ instabilities
in $\Delta _{3}$. The evolution is continuous, without abrupt changes, and
the {\it C}$_{{\it 1}}$ and {\it C}$_{{\it S}}$ instabilities appear rather
as two different levels of deterioration of {\it C}$_{{\it P}}$
instabilities by noise. From this point of view, {\it C}$_{{\it 1}}$
instabilities appear as an intermediate stage between {\it C}$_{{\it P}}$
and {\it C}$_{{\it S}}$ instabilities where noise destroys only the
periodicity, without affecting the cycle itself.

The amplitude of the oscillations, of the order of 100 $\mu $m in $\Delta
_{2}$, is much larger than that of the {\it S} instabilities they merge
from, which is typically 30 $\mu $m. This is an interesting result, because
in the most usual bifurcations between stable and cyclic behaviors, as the
Hopf bifurcation, the cycle merges progressively from a zero amplitude. Such
an atypical behavior can be considered as a signature of the present system,
and must be retrieve in the behavior predicted by the model.

When $\Delta _{0}$ is increased from $\Delta _{2}$, this amplitude still
increases, such that the oscillations reach a 100\% contrast in $z$ and an
amplitude of 40\% in $n$ (Fig. \ref{fig:cs}). Simultaneously, the
instabilities frequency remains almost constant, as illustrated on Fig. \ref%
{fig:VII7} for different values of the intensity $I_{1}$. The frequency
reported here is the main frequency component for {\it C}$_{{\it P}}$ and 
{\it C}$_{{\it 1}}$ instabilities, and the inverse of a mean return time for 
{\it C}$_{{\it S}}$ instabilities. It appears clearly that at given $I_{1}$,
the main frequency does not change between $\Delta _{2}$ and $\Delta _{3}$.
This is another interesting result, also very untypical in dynamical
systems, where the nonlinear resonance frequencies usually depend stricly on
the parameters.

\begin{figure}[tph]
\centerline{\resizebox{0.45\textwidth}{!}{\includegraphics{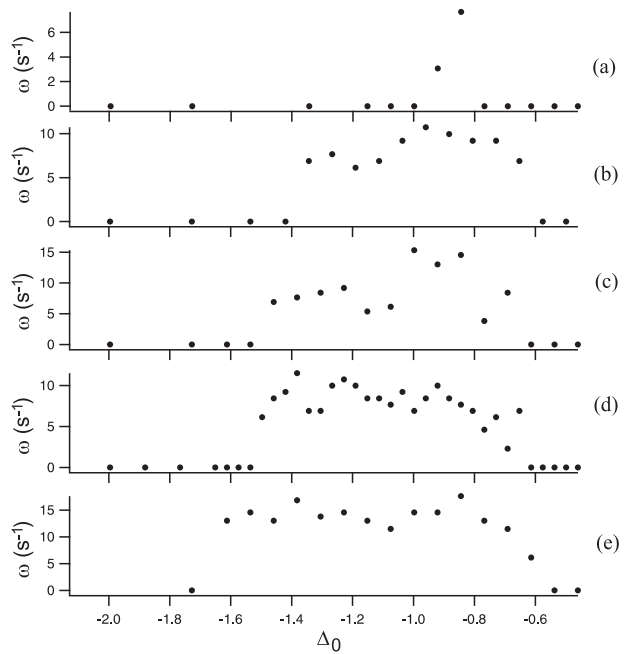}}}
\caption{Experimental evolution of the angular frequency of the
instabilities as a function of the detuning for different values of the beam
internsity: in (a), $I_1=4.1$; in (b), $I_1=6.8$; in (c), $I_1=10$; in (d), $%
I_1=13.6$; in (e), $I_1=18$. $I_rep=1.5$ mW/cm$^2$ in all cases.}
\label{fig:VII7}
\end{figure}

To conclude this section, let us summarize the main characteristics of the 
{\it C} instabilities. Three types of behaviors have been identified,
depending on their degrees of stochasticity: the {\it C}$_{{\it P}}$
instabilities are strictly periodic, the {\it C}$_{{\it 1}}$ instabilities
remain cyclic, but are no more periodic, while finally, the {\it C}$_{{\it S}%
}$ are neither cyclic, nor periodic. However, this last regime differs
drastically from {\it S} instabilities described in \cite{bruit2002}, by
their amplitude, and by a residue of the cycles they merge from. {\it C}
behaviors appear abruptly, with a non-zero amplitude, and their main
frequency appears to be independent from the detuning.

\section{Model}

\label{S4DetMod}To determine the origin and the exact nature of the
instabilities observed in the experiments, we need to build a model able to
reproduce also the complex stochastic dynamics observed in \cite{bruit2002}.
Thus it is logical to use the model introduced in \cite{InstDet} and
described in details in \cite{bruit2002}. It is a 1D model based on the
shadow effect induced by the intensity gradients produced by the absorption
of the trapping laser beams in the cloud \cite{shadow,dalibard}. The aim of
this model is not to reproduce as finely as possible the experimental
system, but on the contrary to be as simple as possible, enlighting the
fundamental mechanisms leading to the instabilities. In its final form, the
model reduces to a set of three autonomous equations, i.e. three equations
not depending explicitely on time. This is important, as it is the minimum
number of degrees of freedom necessary to generate complex dynamics, in
particular deterministic chaos. The model writes: 
\begin{subequations}
\label{eqred}
\begin{eqnarray}
\frac{dZ}{dt} &=&V\frac{v_{r}}{z_{0}}  \label{eqred1} \\
\frac{dV}{dt} &=&\frac{1}{Mv_{r}}F_{T}  \label{eqred2} \\
\frac{dN}{dt} &=&B\left( 1-Z^{2}-N\right)  \label{eqred3}
\end{eqnarray}
\end{subequations}
where $Z=z/z_{0}$, $V=v/v_{r}$\ and $N=n/n_{0}$ are the reduced variables of
the MOT. $z$ and $v$ is the location and the velocity of the center of mass
of the cloud along the unique axis $z$ of the system, while $n$ is the
number of atoms inside the cloud. $z_{0}$ is a phenomenological size
introduced to take into account the transverse distribution of the trap
laser beams, $v_{r}$ is the recoil velocity ($v_{r}=\hbar k/m$), and $n_{0}$
is the equilibrium population of atoms in the cloud. The origin of $z$\
coincides with the \textquotedblleft trap center\textquotedblright , that
is, the zero of the magnetic field. $B$ is the population relaxation rate, $%
M $ the mass of the cloud and $F_{T}$\ the global force exerted on the atoms
by the two counterpropagating beams. To evaluate $F_{T}$, we assume a
multiple scattering regime, i.e. a constant atomic density $\rho $ in the
cloud. Then $F_{T}$ is deduced from the equations of propagation of the
beams inside the cloud\cite{bruit2002}.

Most of the theoretical parameters are the exact counterpart of the
experimental parameters, as e.g. the magnetic field gradient or the beam
intensities. In this case, we used in the model the same values as those of
Table \ref{tabexpparams}. It is not the case for all parameters, either
because of the simplicity of the model or because they cannot be measured
easily in the experiment. In particular, $n_{0}$ and $\rho $ cannot be
accurately evaluated in the experiments. Thus in the simulations, they are
fixed at experimental averaged values, and they have been varied on a wide
range to check their value is not critical. Finally, to perform the
comparison between the experiments and the present model, we sometimes need
to study the behavior of the system when noise is added. This has been done
in the same way as in \cite{bruit2002}, by adding gaussian white noise on $%
I_{1}$. Tab. \ref{tabtheoparams} summarizes the parameters used in the
following.

\begin{table}[h]
\caption{Parameters used in the numerical simulations. The range corresponds
to the interval explored numerically, while the other sets refer to most
of the results presented in this paper.}
\label{tabtheoparams}%
\begin{tabular}{rccc}
& range & set \#1 & set \#2 \\ \hline
$G$ (Gcm$^{-1}$) & $14$ & $14$ & $14$ \\ 
$B$ (s$^{-1}$) & $3\leq B\leq 30$ & $3$ & $3$ \\ 
$I_{1}$ & $2\leq I_{1}\leq 30$ & 25 & 30 \\ 
$\rho $ (cm$^{-3}$) & $10^{10}\leq \rho \leq 3\times 10^{10}$ & $2\times
10^{10}$ & $2\times 10^{10}$ \\ 
$S$ (m$^{2}$) & $10^{-6}\leq S\leq 3\times 10^{-6}$ & $10^{-6}$ & $10^{-6}$
\\ 
$z_{0}$ (m) & $10^{-2}\leq z_{0}\leq 3\times 10^{-1}$ & $3\times 10^{-2}$ & $%
3\times 10^{-2}$ \\ 
$n_{0}$ & $10^{7}\leq n_{0}\leq 10^{9}$ & $10^{8}$ & $6\times 10^{8}$ \\ 
$\Delta _{0}$ & $5\leq \Delta _{0}\leq -0.5$ & $-1.5$ & $-1.5$%
\end{tabular}%
\end{table}

\section{Stationary solutions}

\label{S5StatSol}The model obtained above is described by a set of three
autonomous equations, and thus could exhibit complex behaviors, including
periodic and chaotic oscillations, able to explain the dynamics observed
experimentally. To know if such a complex dynamics occurs effectively with
our parameters, we need first to evaluate the stability of the stationary
solutions, and thus to calculate the stationary solutions themselves. This
work has already been partially presented in \cite{bruit2002}, for stable
stationary solutions, while we are interested here by unstable stationary
solutions. However, for sake of clarity, we recall here some of the general
results given in \cite{bruit2002}, before to start the analysis of the
unstable stationary solutions.

\begin{figure}[tph]
\centerline{\resizebox{0.45\textwidth}{!}{\includegraphics{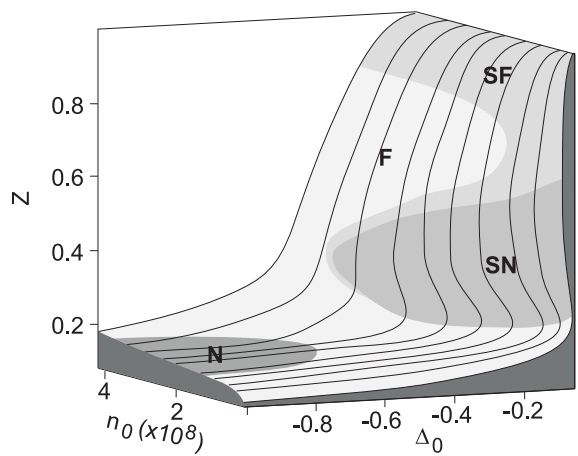}}}
\caption{Stationary solutions of equations \protect\ref{eqred} versus $n_{0}$
and $\Delta _{0}$. The figure represents $Z_{s}$. Other parameters
correspond to the set \#1 given in table \protect\ref{tabtheoparams}. N, F,
SN and SF zones (each corresponding to different level of greys) describes
the nature of the fixed point associated with the stationary solution:
stable Node, stable Focus, Saddle Node and Saddle Focus. }
\label{fig:ZS}
\end{figure}

\begin{figure}[tph]
\centerline{\resizebox{0.45\textwidth}{!}{\includegraphics{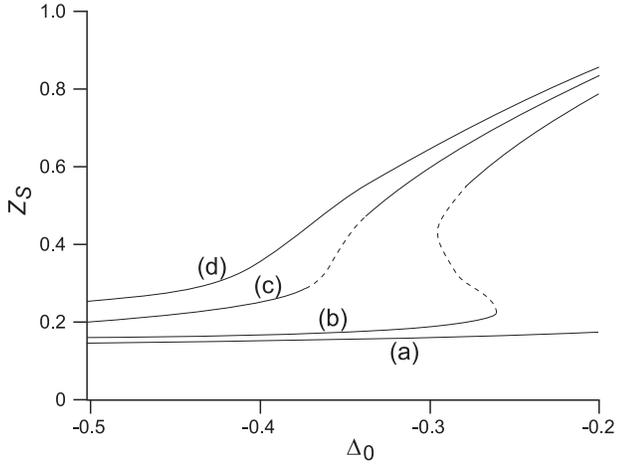}}}
\caption{Evolution as a function of the detuning of the stationary solution $%
Z_s$ of equations \protect\ref{eqred}. The full (resp. dashed) line
corresponds to a stable (resp. unstable) solution. In (a), $n_0=0.5\times
10^8$; in (b) $n_0=2.5\times 10^8$; in (c), $n_0=3.4\times 10^8$; in (d), $%
n_0=4\times 10^8$ Other parameters correspond to the set \#1 of Table 
\protect\ref{tabtheoparams}}
\label{fig:foldex}
\end{figure}

The three stationary solutions $Z_{s}$, $V_{s}$ and $N_{s}$ are given by Eq.%
\ref{eqred}, when the left sides are put to zero. As discussed in \cite%
{bruit2002}, $V_{s}$ and $N_{s}$ can be deduced easily from $Z_{s}$, and
thus the discussion is reduced to that of $Z_{s}$, the equation of which can
be resolved numerically. The global shape of $Z_{s}$ is illustrated in Fig. %
\ref{fig:ZS}, where it is plotted as a function of $\Delta _{0}$\ and $n_{0}$%
. The basic characteristic of this diagram is the fold in the stationary
solutions, due to several abrupt slope changes. The shape of the fold
depends on the parameters, in particular on $n_{0}$. Fig. \ref{fig:foldex}
shows four examples corresponding to a situation leading to basically
different atomic dynamics. For $n_{0}=0.5\times 10^{8}$ (fig. \ref%
{fig:foldex}a), $Z_{s}$ increases smoothly with $\Delta _{0}$ (i.e. $N_{s}$
decreases slowly). For $\Delta _{0}\simeq 0.1$ (and thus outside the graph),
the cloud vanishes through a narrow bistable cycle, where $Z_{s}$ jumps
abruptly from a value of the order of $0.2$ to a value close to $1$. As $%
n_{0}$ increases, this bistable cycle appears for smaller $Z_{s}$ (and thus
larger $N_{s}$), and becomes physically significant. Fig. \ref{fig:foldex}b
shows $Z_{s}$ for $n_{0}$\ $=2.5\times 10^{8}$ and a bistable cycle for $%
-0.3\lesssim \Delta _{0}\lesssim -0.25$. If $n_{0}$ is further increased,
the bistable cycle disappears, but it remains a fold corresponding to two
abrupt slope changes of $Z_{s}$ versus $\Delta _{0}$ (fig. \ref{fig:foldex}%
c, $n_{0}$\ $=3.4\times 10^{8}$). If $n_{0}$ is still increased, the fold
remains, but it becomes smoother (Fig. \ref{fig:foldex}d for $n_{0}=$\ $%
4\times 10^{8}$).

The results of the linear stability analysis have been detailed in \cite%
{bruit2002}. It was shown that the stability and nature of the stationary
solutions evolve along the fold. In particular, the solutions are unstable
not only on the central branch of the bistable cycle, as it is usual, but
also on the upper branch of the bistable cycle, and even when there is no
bistability. This is illustrated in Fig. \ref{fig:foldex}, where the
unstable solutions are plotted in a dashed line. As we deal here with
deterministic instabilities, the interesting situation corresponds to Fig. %
\ref{fig:foldex}c, where the unique stationary solution is unstable on the
fold. Note the difference with the cases studied in \cite{bruit2002}, where
the stationary solution is also unique, but stable. In the present case, as
no stable solution exists, the system exhibits necessarily deterministic
instabilities.

\begin{figure}[tph]
\centerline{\resizebox{0.45\textwidth}{!}{\includegraphics{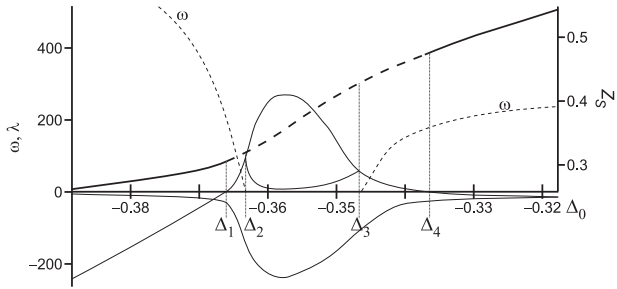}}}
\caption{Evolution as a function of the detuning of the stationary solution $%
Z_S$ and its eigenvalues, for the parameters of Fig. \protect\ref{fig:foldex}%
c. The stationary solution is given through the full (stable) and dashed
(unstable) bold lines. The dashed line noted $\protect\omega$ represents the
imaginary part of the complex eigenvalues, while the full lines correspond
to the real part of the three eigenvalues. The full line remaining negative
corresponds to the eigenvalue which is real everywhere.}
\label{fig:foldinst}
\end{figure}

Fig. \ref{fig:foldinst} details the changes in the eigenvalues in this
situation. Outside the fold (i.e. $\Delta _{0}<\Delta _{1}$ or $\Delta
_{0}>\Delta _{4}$), $Z_{S}$ is stable, with one real eigenvalue $\lambda _{r}
$ and two complex conjugate eigenvalues $\lambda \pm i\omega $: the fixed
point associated to the stationary solution in the phase space is a stable
focus (F zone in Fig. \ref{fig:ZS}). The real numbers $-\lambda _{r}$ and $%
-\lambda $ are the damping rates of the stationary solution, and $\omega $
its eigenfrequency. The transition from stable focus solution to unstable
saddle focus solution occurs in $\Delta _{1}$ and $\Delta _{4}$, through a
Hopf bifurcation, where $\lambda =0$ and $\omega \neq 0$. As it is usual for
such a bifurcation, we expect to observe, on the unstable side, a cloud
moving on a limit cycle with a frequency $\omega $, i.e. in the present case
of the order of 200 s$^{-1}$ (30 Hz) for both Hopf bifurcations. Another
transition, from a saddle focus solution to a saddle node solution, occurs
in $\Delta _{2}$ and $\Delta _{3}$, when $\omega $ vanishes. The behavior
here cannot be deduced from the stationary solutions, and numerical
simulations are necessary. Deterministic instabilities are expected to occur
between $\Delta _{1}$ and $\Delta _{4}$. In all other areas, the stationary
solutions are stable, and therefore, deterministic instabilities cannot occur%
\cite{bruit2002}. In the next section, we detail the results of numerical
simulations performed in the unstable area.

However, before to discuss in detail the dynamical behaviors predicted by
the model in the different situations, let us look at the influence of the
other parameters on the stationary solutions. As discussed in the previous
section, we must distinguish between theoretical parameters with an exact
experimental counterpart, as $I_{1}$, for which the comparison with
experiments is direct, from those without an exact experimental counterpart,
for which the analysis is more delicate. It is in particular the case for $%
n_{0}$ and $\rho $, which are both linked to $I_{rep}$ and the vapour
pressure. Finally, the influence of $B$ and $z_{0}$ should be checked, as
their experimental determination is unprecise.

\begin{figure}[tph]
\centerline{\resizebox{0.45\textwidth}{!}{\includegraphics{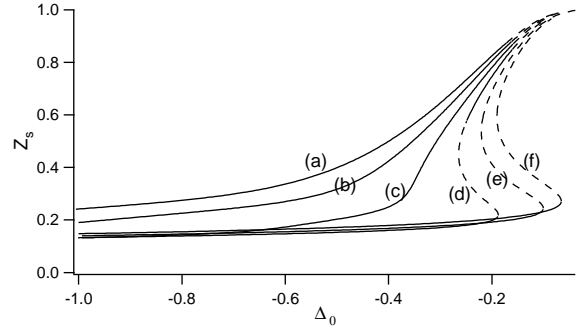}}}
\caption{Evolution as a function of the detuning of the stationary solution $%
Z_s$ of equations \protect\ref{eqred} for different values of the intensity $%
I_1$. The full (resp. dashed) line corresponds to a stable (resp. unstable)
solution. In (a), $I=10$; in (b) $I=15$; in (c), $I=20$; in (d), $I=25$; in
(e), $I=30$; in (d), $I=35$. Other parameters correspond to the set \#1 of
Table \protect\ref{tabtheoparams}, with $n_0=2\times 10^8$.}
\label{fig:IVar}
\end{figure}

\begin{figure}[tph]
\centerline{\resizebox{0.45\textwidth}{!}{\includegraphics{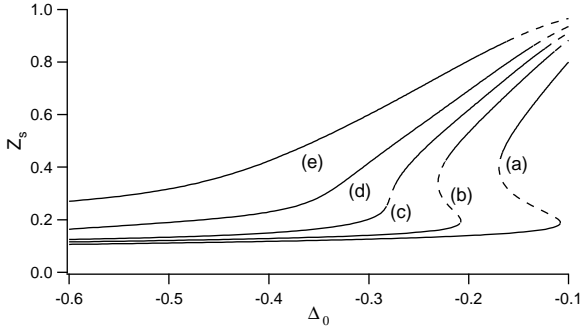}}}
\caption{Evolution as a function of the detuning of the stationary solution $%
Z_s$ of equations \protect\ref{eqred} for different values of $n_0$ and a
smaller intensity than in Fig. \protect\ref{fig:foldex}. The full (resp.
dashed) line corresponds to a stable (resp. unstable) solution. In (a), $%
n_0=0.25\times 10^8$; in (b) $n_0=0.5\times 10^8$; in (c), $n_0=0.7\times
10^8$; in (d), $n_0=1\times 10^8$; in (e), $n_0=2\times 10^8$. Other
parameters correspond to the set \#1 of Table \protect\ref{tabtheoparams},
with $I_1=15$.}
\label{fig:ZSI}
\end{figure}

Fig. \ref{fig:IVar} illustrates the role of the $I_{1}$ value on the
behavior of the cloud: it represents the stationary solution $Z_{S}$ versus
the detuning, for different values of the intensity. The figure shows that $%
I_{1}$ acts as $n_{0}$ on $Z_{S}$: an increase of $I_{1}$ makes the fold
steeper, and eventually leads to bistability. However, some more subtle
changes occur, as illustrated by Fig. \ref{fig:ZSI}, where $Z_{S}$ have been
plotted as a function of $\Delta _{0}$ for different values of $n_{0}$, as
in Fig. \ref{fig:foldex}, but for a smaller intensity. In these new
conditions, the intermediate area between the stable fold and bistability,
where the stationary solution is unique and unstable, has almost disappear.
This result may be generalized: in the simulations, we observed that the
area corresponding to an unstable unique solution disappears for small
intensities (typically $I_{1}<10$). This result is in agreement with the
experimental observation that {\it C} instabilities exist only for large
intensities.

\begin{figure}[tph]
\centerline{\resizebox{0.45\textwidth}{!}{\includegraphics{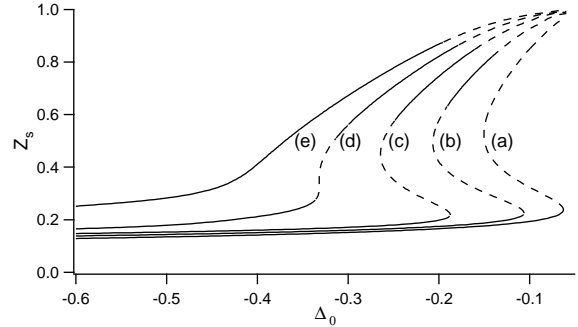}}}
\caption{Evolution as a function of the detuning of the stationary solution $%
Z_s$ of equations \protect\ref{eqred} for different values of the atomic
density. The full (resp. dashed) line corresponds to a stable (resp.
unstable) solution. In (a), $\protect\rho=1\times 10^{10}$ cm$^{-3} $; in
(b), $\protect\rho=1.5\times 10^{10}$ cm$^{-3}$; in (c), $\protect\rho%
=2\times 10^{10}$ cm$^{-3}$; in (d), $\protect\rho=2.5\times 10^{10}$ cm$%
^{-3}$; in (e), $\protect\rho=3\times 10^{10}$ cm$^{-3}$. Other parameters
correspond to the set \#1 of Table \protect\ref{tabtheoparams}, with $%
n_0=2\times 10^8$ and $I_1=25 $.}
\label{fig:roVar}
\end{figure}

\begin{figure}[tph]
\centerline{\resizebox{0.45\textwidth}{!}{\includegraphics{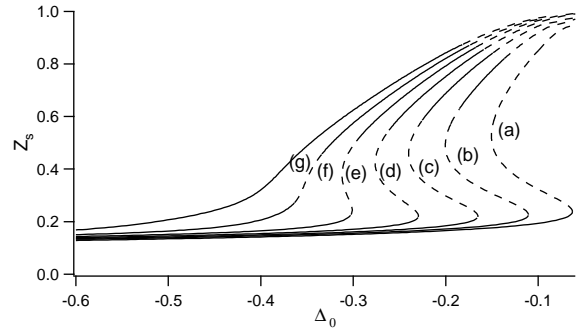}}}
\caption{Evolution as a function of the detuning of the stationary solution $%
Z_s$ of equations \protect\ref{eqred} for different values of $n_0$ and a
smaller atomic density than in Fig. \protect\ref{fig:foldex}. The full
(resp. dashed) line corresponds to a stable (resp. unstable) solution. In
(a), $n_0=2\times 10^8$; in (b) $n_0=4\times 10^8$; in (c), $n_0=6\times
10^8 $; in (d), $n_0=8\times 10^8$; in (e), $n_0=1\times 10^9$; in (f), $%
n_0=1.2\times 10^9$; in (g), $n_0=1.4\times 10^9$. Other parameters
correspond to the set \#1 of Table \protect\ref{tabtheoparams}, with $%
\protect\rho=1\times 10^{10}$ cm$^{-3}$.}
\label{fig:ZSro}
\end{figure}

The atomic density value used in the present simulation is $\rho =2\times
10^{10}%
\mathop{\rm cm}%
^{-3}$, which represents an average of the density we measured
experimentally. To evaluate the influence of this value on the predicted
behavior, we have plotted in Fig. \ref{fig:roVar} the evolution of $Z_{S}$
for different values of $\rho $. As for $n_{0}$ and $I_{1}$, the $Z_{S}$
curve evolves from an almost flat dependence for large $\rho $, towards
bistability for small $\rho $, with an intermediate SF zone. This can appear
as surprising, because it seems to mean that nonlinear behaviors,
corresponding to the bistable cycle, need a small atomic density. In fact,
this reasoning is false, because it does not take into account the role of
the other parameters, in particular $n_{0}$, which is able to compensate for
the variation of $\rho $. For example, fig. \ref{fig:ZSro} shows the $\left(
\Delta _{0},n_{0}\right) $ diagram for a smaller $\rho $ value than in Fig. %
\ref{fig:foldex}: it has the same properties as that in fig. \ref{fig:foldex}%
, except that the population are much larger. However, one remarks slight
differences, in particular a small decreasing of the $Z$ values, and first
of all a small decreasing of the SF zone width. This is another general
result: in the simulations, when the atomic density is decreased, the $Z_{s}$
curves globally flatten, so that the fold becomes less steep. Thus all
instabilities disappear for very small densities. This is in agreement with
the experimental results illustrated in Fig. \ref{fig:largeurvsI}b, where
the unstable interval width is reported as a function of $I_{rep}$: the
instabilities disappear for small $I_{rep}$, i.e. when the efficiency of the
repumper -- and thus the atomic density -- decreases.

\begin{figure}[tph]
\centerline{\resizebox{0.45\textwidth}{!}{\includegraphics{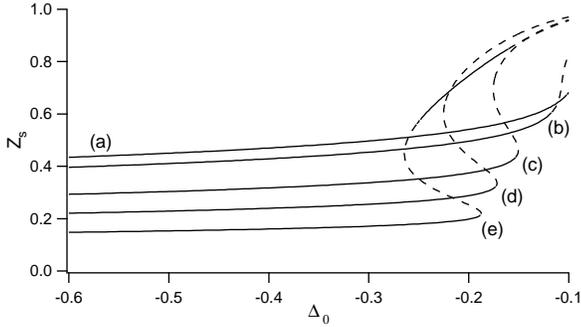}}}
\caption{Evolution as a function of the detuning of the stationary solution $%
Z_s$ of equations \protect\ref{eqred} for different values of $z_0$. The
full (resp. dashed) line corresponds to a stable (resp. unstable) solution.
In (a), $z_0=1$ cm; in (b), $z_0=1.1$ cm; in (c), $z_0=1.5$ cm; in (d), $%
z_0=2$ cm; in (e), $z_0=3$ cm. Other parameters correspond to the set \#1 of
Table \protect\ref{tabtheoparams}, with $n_0=2\times 10^8$ and $I_1=25$.}
\label{fig:z0Var}
\end{figure}

As detailed in \cite{bruit2002}, the value of $z_{0}=3%
\mathop{\rm cm}%
$ used in the experiments has been evaluated from the trap beam waist,
taking into account the beam intensity as compared to the saturation
intensity. As for $\rho $, we want to evaluate how critical is this choice
by plotting $Z_{S}$ versus $\Delta _{0}$ for different values of $z_{0}$
(Fig. \ref{fig:z0Var}). It appears clearly that a decreasing of $z_{0}$
corresponds to a shift of the fold and the bistable cycle towards resonance.
Thus, for $z_{0}$ values smaller than 3 cm, the discrepancy between
simulations and experiments increases quantitatively, as instabilities will
appear at smaller detuning. For values smaller than 1 cm, the change is
drastic, as the SF and bistable zones, and thus instabilities, disappear. On
the contrary, for larger $z_{0}$, the bistable cycle widens and shifts off
resonance.

The last parameter to be considered is $B$. This parameter appears to be not
critical at all, and in particular, a change from e.g. $B=5%
\mathop{\rm s}%
^{-1}$ to $B=1%
\mathop{\rm s}%
^{-1}$ does not change in a noticeable way the values of $Z_{S}$. The main
change concerns the smaller real eigenvalue, which takes typically the value
of $-B$. However, this eigenvalue plays a minor role in the dynamics, as it
remains always real negative, and thus this change has a negligible effect
on the dynamics.

It appears from the above analysis that the existence of the unstable area
does not depend critically on the values of $n_{0}$, $I_{1}$, $\rho $, $%
z_{0} $ and $B$. In particular, the relative poor accuracy in the knowledge
of the experimental values of some parameters, as $n_{0}$, $\rho $ or $z_{0}$%
, does not appear as a limitation in the above study, because a change of
some tens of percents around the default values used in the simulations does
not alter the results. The numerous approximations at the origin of the
model lead probably to larger errors.

\section{The unstable fold: deterministic instabilities}

\label{S6Dyn}
\begin{figure}[tph]
\centerline{\resizebox{0.45\textwidth}{!}{\includegraphics{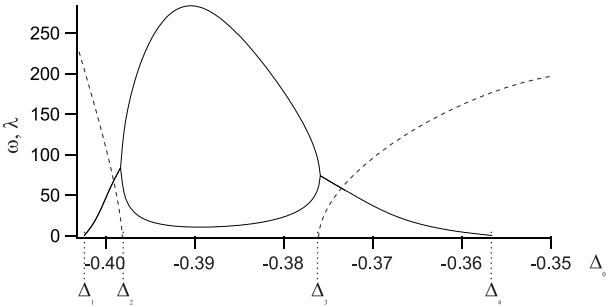}}}
\caption{Evolution as a function of the detuning of the eigenvalues of the
stationary solutions. The full line, which is a plot of the real part $%
\protect\lambda$ of the eigenvalues when they are positive, put in evidence
two bifurcations, in $\Delta_1=-0.4024$ and $\Delta_4=-0.3566$. The dashed
line represents the corresponding imaginary part $\protect\omega$ (i.e. the
eigenfrequencies). The eigenvalues are real between $\Delta_2=-0.3981$ and $%
\Delta_3=-0.3760$. Parameters corresponds to the set \#2 of Tab. \protect\ref%
{tabtheoparams}.}
\label{fig:FI}
\end{figure}

As shown in the previous section, it exists a range of
parameters where the stationary solution is unique and unstable. Such a
situation leads inevitably to deterministic instabilities, with shape and
characteristics obtained through numerical simulations of the model for the
corresponding sets of parameters. In the present section, we discuss the
behavior obtained by such simulations, in a situation similar to Figs \ref%
{fig:foldex}c and \ref{fig:foldinst}, but for a slightly different set of
parameters (set \#2 of Tab. \ref{tabtheoparams}). Fig. \ref{fig:FI} shows
for these conditions the evolution of the eigenvalues as a function of the
detuning. The sequence is identical to that followed in Fig. \ref%
{fig:foldinst}, but the unstable zone is wider. Starting off resonance, a
Hopf bifurcation occurs in $\Delta _{1}$. In $\Delta _{2}$, the eigenvalues
become real, and thus the eigenfrequency disappears, until $\Delta _{3}$,
where two eigenvalues are again complex. Finally, a second Hopf bifurcation
occurs in $\Delta _{4}$, and the stationary solutions become stable again.

\subsection{The behavior in the vicinity of the Hopf bifurcation}

To understand the origin of the {\it C} instabilities, we have studied in
detail the behavior of the model in the close vicinity of the H$_{1}$ Hopf
bifurcation, by varying $\Delta _{0}$ slightly above $\Delta _{1}$. The
expected scenario has been extensively described in the litterature\cite{RT}%
. On the left of H$_{1}$,\ the stationary solution is stable, and thus the
behavior is stationary. In H$_{1}$, the solution becomes unstable, but a
limit cycle merges from the unstable fixed point: the behavior becomes
periodic, with a zero amplitude in H$_{1}$, and a frequency corresponding to
the relaxation frequency of the unstable solution. On the right of the
bifurcation, the amplitude of the oscillations grows, while the frequency
remains the same as the eigenfrequency in the vicinity of H$_{1}$. Usually,
when the control parameter is increased from H$_{1}$, the cycle shape
changes progressively, while the interval between the oscillation and
relaxation frequencies becomes larger. This standard scenario is absolutely
not followed by the present model. On the contrary, several abrupt changes
in the behavior occur in a very narrow interval of $\Delta _{0}$, leading
from a classic regular limit cycle to the {\it C}$_{P}$ instabilities.

\begin{figure}[tph]
\centerline{\resizebox{0.45\textwidth}{!}{\includegraphics{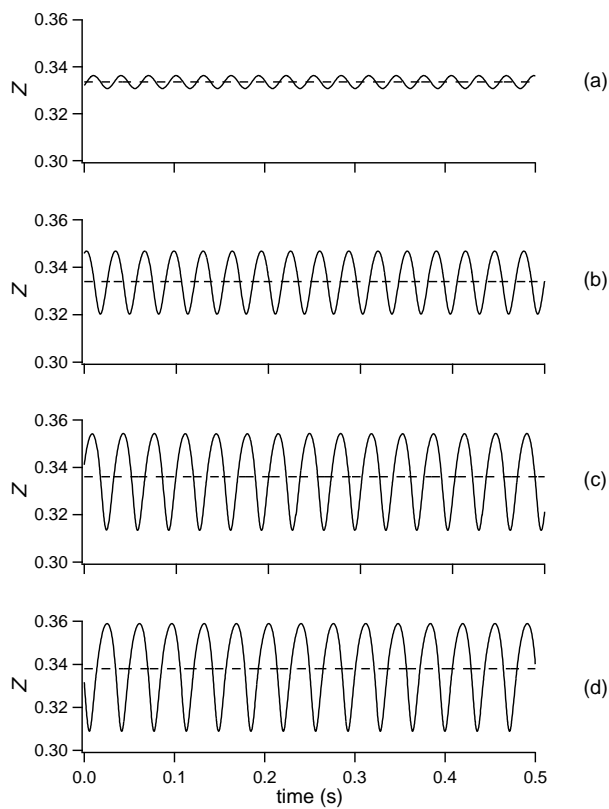}}}
\caption{Time evolution of the atomic cloud location $Z$ in the close
vicinity of the $\Delta _{1}$ bifurcation, on the unstable side. The dashed
line represents the value of the unstable stationary solution $Z_{S}$. In
(a), $\Delta _{0}=-0.4023$; in (b), $\Delta _{0}=-0.4020$; in (c), $\Delta
_{0}=-0.4015$; in (d), $\Delta _{0}=-0.4010$. Other parameters are those of
set \#2 of Tab. \protect\ref{tabtheoparams}.}
\label{fig:zper}
\end{figure}

In H$_{1}$ appears, as expected, a limit cycle. Figs. \ref{fig:zper} and \ref%
{fig:nper} show the evolution of this limit cycle on the unstable side of
the bifurcation between $\Delta _{1}$ and $\Delta _{1}^{\prime }=-0.4010$.
Note that the explored interval is so narrow ($1.3\times 10^{-3}$) that an
experimental observation of the described phenomena cannot be considered.
Following the $Z$ coordinate, the cycle amplitude grows rapidly to reach a
value of typically $10\%$ of $Z_{S}$, while the amplitude on $N$ remains
very small ($0.1\%$ of $N_{S}$). The cycle remains relatively well centered
on $Z_{S}$, but is shifted compared to $N_{S}$, such that $N_{S}$ is well
outside the cycle. This is not an exceptional situation, as it simply means
that the basin of attraction of the cycle is curved in the vicinity of the
unstable fixed point. Another characteristics of the limit cycle is its
frequency, which remains of the order of magnitude of the eigenfrequency.
For example, for $\Delta _{0}=-0.4015$ (Fig. \ref{fig:zper}c), the behavior
frequency is 30 Hz, for an eigenvalue of 27 Hz. Thus the global behavior in
the interval $(\Delta _{1},\Delta _{1}^{\prime })$ appears to be the usual
one in the vicinity of a Hopf bifurcation.

\begin{figure}[tph]
\centerline{\resizebox{0.45\textwidth}{!}{\includegraphics{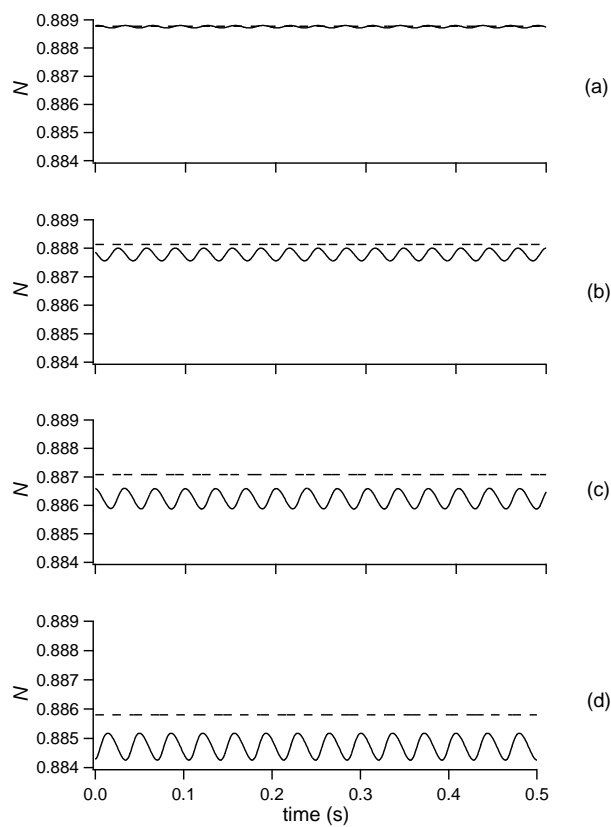}}}
\caption{Time evolution of the atomic cloud population $N$ in the close
vicinity of the $\Delta _{1}$ bifurcation, on the unstable side. The dashed
line represents the value of the unstable stationary solution $N_{S}$.
Parameters are the same as in Fig \protect\ref{fig:zper}.}
\label{fig:nper}
\end{figure}

\begin{figure}[tph]
\centerline{\resizebox{0.45\textwidth}{!}{\includegraphics{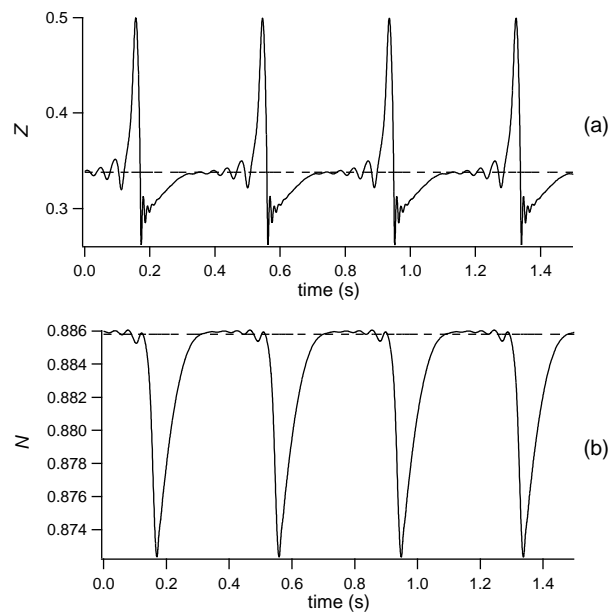}}}
\caption{Time evolution of the atomic cloud location $Z$ in (a), and
population $N$ in (b), for $\Delta _{0}=-0.4010$. Other parameters are the
same as set \#2 in Tab. \protect\ref{tabtheoparams}.}
\label{fig:homo}
\end{figure}

However, for $\Delta \geq \Delta _{1}^{\prime }$, the limit cycle becomes
unstable and is replaced by another periodic orbit, with a much more complex
shape (Fig. \ref{fig:homo}) and a much longer period. The amplitude is
almost 5 times larger for $Z$ and more than 10 times for $N$. The trajectory
consists in several different stages: a diverging spiral off the fixed
point, followed by a large loop and a convergent spiral until the fixed
point. The frequencies of the two oscillating stages are different: this is
not surprising, as the diverging one is clearly linked to the fixed point,
and thus to its eigenfrequency, contrary to the convergent one. One finds
effectively a frequency of 24.4 Hz for the diverging spiral, corresponding
exactly to the eigenfrequency ($\omega =2\pi \times 23.5%
\mathop{\rm Hz}%
$), while the frequency of the converging spiral is 77 Hz. However, the main
frequency of the behavior is 2.6 Hz, i.e. one order of magnitude slower than
that of the Hopf cycle. The properties of the trajectories, in particular
the tangency to the unstable point and the large loop in the phase space,
are characteristic from a homoclinic behavior, when the stable and unstable
manifolds of the fixed point are almost connected. A accurate analysis of
these manifolds would be necessary to conclude about this point. Note that
in a very narrow interval around $\Delta _{1}^{\prime }$, generalized
bistability occurs between the Hopf cycle and the homoclinic one.

\begin{figure}[tph]
\centerline{\resizebox{0.45\textwidth}{!}{\includegraphics{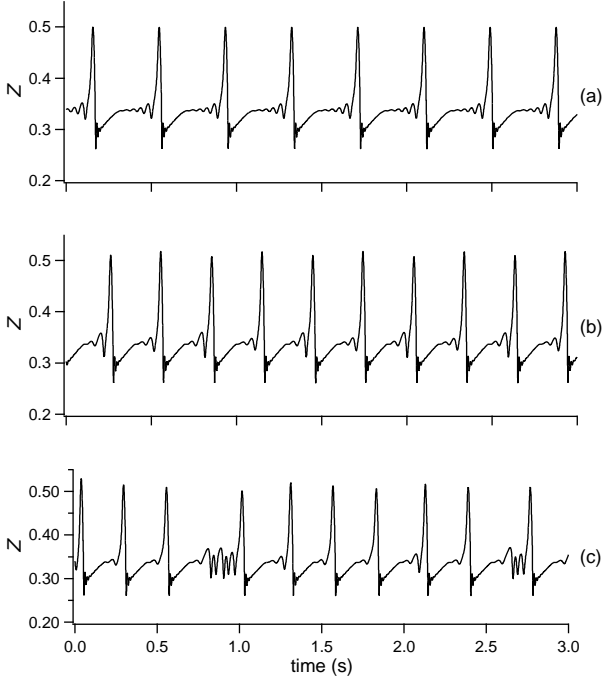}}}
\caption{Time evolution of the atomic cloud location $Z$ in the vicinity of
the chaotic zone. In (a), for $\Delta _{0}=-0.4010$, the motion is periodic;
in (b), for $\Delta _{0}=-0.4006$, period doubling appears; in (c), for $%
\Delta _{0}=-0.4005$, the motion is chaotic. Other parameters are the same
as set \#2 in Tab. \protect\ref{tabtheoparams}.}
\label{fig:chaos}
\end{figure}

\begin{figure}[tph]
\centerline{\resizebox{0.45\textwidth}{!}{\includegraphics{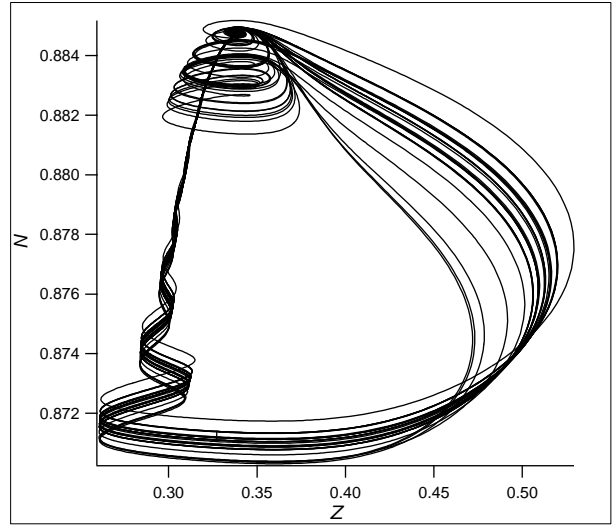}}}
\caption{Chaotic trajectories of the cloud motion in the $(Z,N)$ phase
space. $\Delta_0=-0.4005$. Other parameters are the same as in Tab.  \protect
\ref{tabtheoparams}.}
\label{fig:attracteur}
\end{figure}

When $\Delta _{0}$ is increased from $\Delta _{1}^{\prime }$, the cloud
exhibits period doubling and chaos (Fig. \ref{fig:chaos}). The trajectories
keep the same shape, in particular with the two spiraling episodes and the
large loop, but the periodicity is modified or is lost. For example, when
the period is double, variations appear essentially on the amplitude of the
loop together with that of the diverging oscillations (Fig. \ref{fig:chaos}%
b). In the chaotic zone, the irregularities appear also on these amplitudes,
but bursting events appear sometimes between these two stages (Fig. \ref%
{fig:chaos}c). We did not perform a precise analysis of these behaviors,
mainly because they appear on a so narrow interval that there is no chance
to observe them experimentally. Indeed, chaos disappears for $\Delta >\Delta
_{1}^{\prime \prime }$, with $\Delta _{1}^{\prime \prime }=-0.4005$, and
thus the homoclinic behavior appears on an interval of $5\times 10^{-4}$.
However, a simple test can be done by reconstructing the attractor of the
dynamics (Fig. \ref{fig:attracteur}). A glance at the result shows a
definite structure, and not random distributed points, and thus confirms
that this behavior presents all the characteristics of deterministic chaos.

\subsection{$C_{P}$ instabilities}

\begin{figure}[tph]
\centerline{\resizebox{0.45\textwidth}{!}{\includegraphics{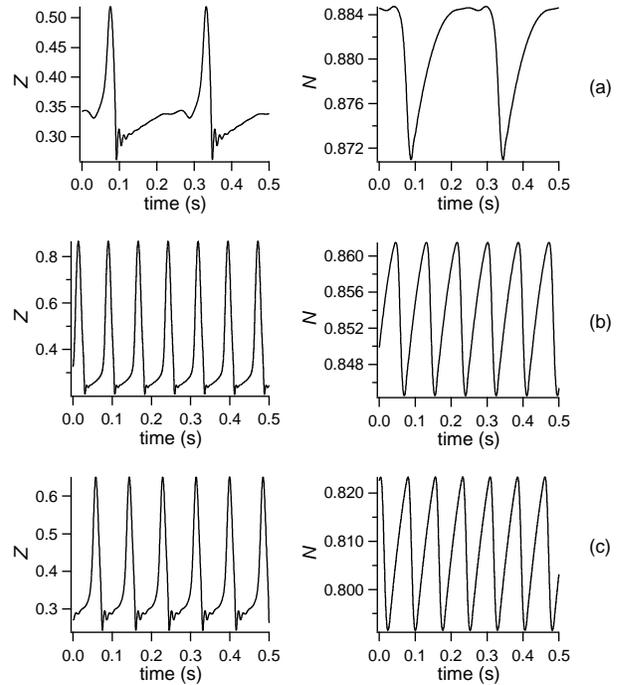}}}
\caption{Time evolution of the atomic cloud location $Z$ (left) and
population $N$ (right) in the unstable zone. In (a), $\Delta_0=-0.4004$; in
(b), $\Delta_0=-0.39$; in (c), $\Delta_0=-0.37$. Other parameters are the
same as in Tab. \protect\ref{tabtheoparams}.}
\label{fig:CPt}
\end{figure}

For $\Delta _{0}>\Delta _{1}^{\prime \prime }$, the homoclinic behavior
disappears, and a new type of periodic instabilities appear (Fig. \ref%
{fig:CPt}). There is no fundamental difference between the homoclinic
instabilities and the present behavior, except that the latter has a
physical meaning, as it appears in the simulations on a significant $\Delta
_{0}$ interval, about $0.04\Gamma $ for the present parameters. One observes
in Fig. \ref{fig:CPt}a the same three stages as in Fig. \ref{fig:chaos},
which means that the origin of this behavior is the same as for the
homoclinic instabilities. However, these three stages exist only in the
close vicinity of $\Delta _{1}^{\prime \prime }$: when $\Delta _{0}$ is
increased, the diverging helix around the fixed point disappears, and only
the two stages independent from the fixed point remain (Fig. \ref{fig:CPt}b
and c). This means that in this new behavior, the trajectories never
approach the fixed point, and thus the dynamics does not depend on the local
properties of the fixed point.

Each period of the new behavior may be divided in two stages with different
durations. During the fast stage, $Z$ makes an oscillation, first growing
then decreasing, while $N$ decreases; during the slow stage, $Z$ and $N$
grow. This is similar to that observed in experiments with {\it C}$_{{\it P}%
} $ instabilities, and thus it is interesting to check if the other
properties of {\it C}$_{{\it P}}$ instabilities can be retrieve in the
present dynamics.

\begin{figure}[tph]
\centerline{\resizebox{0.45\textwidth}{!}{\includegraphics{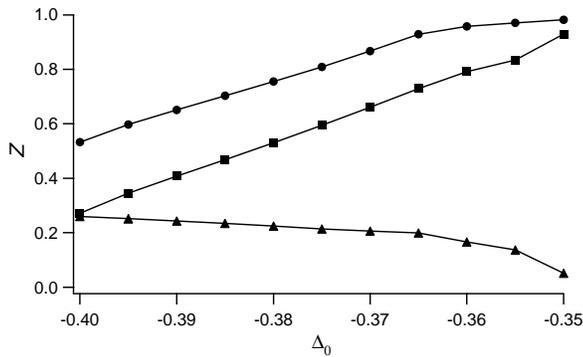}}}
\caption{Amplitude of the dynamics versus detuning. The squares, discs and
triangles corresponds respectively to the oscillation amplitude, the maximum
value and the minimum value reached by $z$. Parameters are those of Tab. 
\protect\ref{tabtheoparams}.}
\label{fig:Amplitude}
\end{figure}

\begin{figure}[tph]
\centerline{\resizebox{0.45\textwidth}{!}{\includegraphics{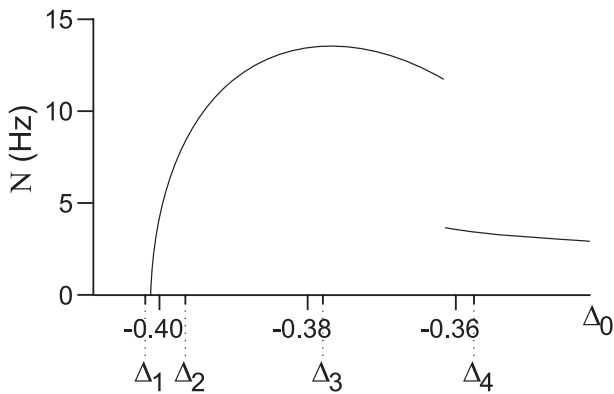}}}
\caption{Evolution as a function of the detuning of the instabilities
frequency.}
\label{fig:freq}
\end{figure}

\begin{figure}[tph]
\centerline{\resizebox{0.45\textwidth}{!}{\includegraphics{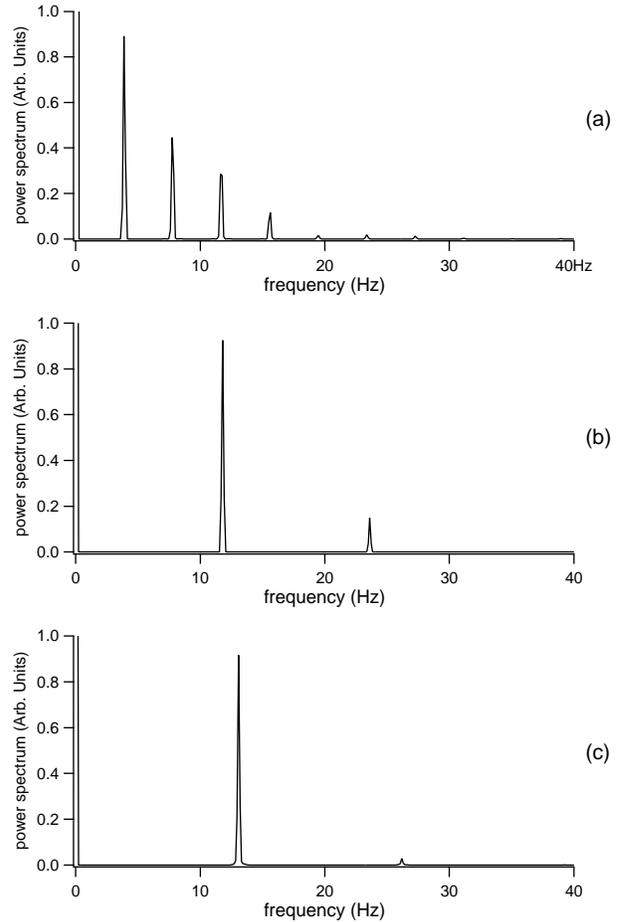}}}
\caption{Power spectra of the atomic cloud location $Z$ in the unstable
zone. Parameters are the same as in Fig.\protect\ref{fig:CPt}.}
\label{fig:CPfft}
\end{figure}

Two untypical properties were noticed in the {\it C}$_{{\it P}}$
instabilities, concerning the non zero amplitude of the oscillations when
they appear, and their almost constant frequency along their interval of
existence. The amplitude of the oscillations as a function of $\Delta _{0}$
is plotted in fig. \ref{fig:Amplitude}. In $\Delta _{0}=-0.4$, when the
instabilities appear, their amplitude is already almost 0.3. In fact,
because the Hopf limit cycle exists on a very narrow interval, this is not
strictly true. But from a physical point of view, it is clear that
instabilities appear with a non zero amplitude, as in the experiments.
Concerning the frequency, its value tends to zero in the vicinity of $\Delta
_{1}^{\prime \prime }$, but it increases rapidly to reach a value of the
order of 10 Hz, and remains between 10 Hz and 13 Hz on most of the unstable
interval, as shown in Fig. \ref{fig:freq} (the behavior for $\Delta
_{0}>-0.36$ is discussed below). Finally, to complete the comparison with
the experiments, Fig. \ref{fig:CPfft} shows the power spectra of $Z$. In the
vicinity of H$_{1}$, it is very characteristic, with a main frequency and
large amplitude harmonics decreasing progressively (Fig. \ref{fig:CPfft}a),
as in the experiments (Fig. \ref{fig:VII5}). As the detuning is increased,
the amplitude of the harmonics decreases, and several new frequencies appear
in the spectrum, but each component has individually an amplitude negligible
compared to the main frequency (Fig. \ref{fig:CPfft}b and c).

To summarize, we found that the periodic instabilities in the vicinity of H$%
_{1}$ have the same shape, spectrum, amplitude evolution and frequency
evolution as the {\it C}$_{{\it P}}$ instabilities observed in the
experiments; all these points confirm that the model reproduces here the 
{\it C}$_{{\it P}}$ instabilities (Fig. \ref{fig:cp} and \ref{fig:CPt}).

However, several quantitative discrepancies appear between the present
results and the experimental observations, as e.g. for the $\Delta _{0}$
values or the instabilities frequency. These differences are relatively
small, except for the $\Delta _{0}$ interval where {\it C}$_{{\it P}}$
instabilities exist: it is typically of the order of $1$ in the experiments,
while it is smaller than $0.1$ in the simulations. This last value could be
increased by increasing the value of $I_{1}$ in the simulations, reducing
the difference to less than a factor 10. Considering the extreme simplicity
of our model and its numerous approximations, it is clear that it is able to
reproduce strikingly the {\it C}$_{{\it P}}$ instabilities, with a
surprisingly good agreement.

\subsection{{\it C}$_{{\it 1}}$ instabilities}

In the experiments, {\it C}$_{{\it P}}$ instabilities are replaced, as the
detuning is changed, by {\it C}$_{{\it 1}}$ instabilities. The difference
between the two regimes appears mainly on the $n$ behavior, and in
particular in the shape of the signal, which loses its regularity. The {\it C%
}$_{{\it 1}}$ behavior is not observed in the present model. However, as
discussed in the experimental section, the {\it C}$_{{\it 1}}$ instabilities
do not appear as a new deterministic regime, but rather as {\it C}$_{{\it P}%
} $ instabilities slightly altered by noise. Thus, to test the ability of
the present model to reproduce this behavior, it is necessary to add noise
in the model. The results obtained in this case are discussed in the next
section.

\subsection{$C_{S}$ instabilities}

\begin{figure}[tph]
\centerline{\resizebox{0.45\textwidth}{!}{\includegraphics{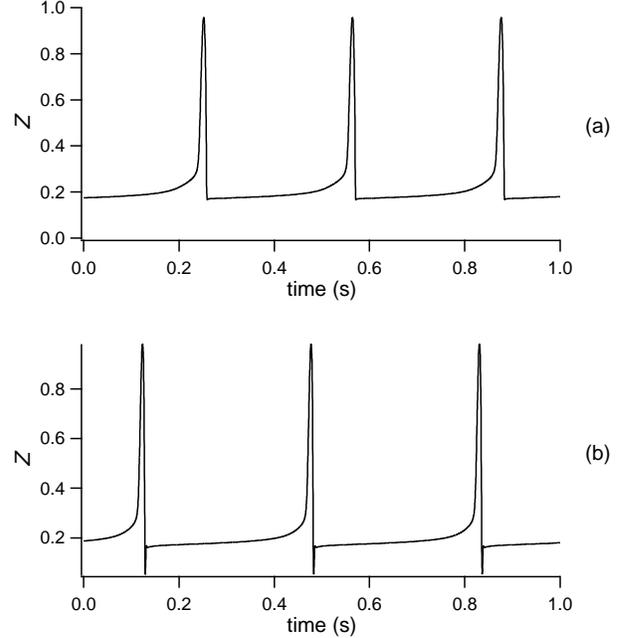}}}
\caption{Time evolution of the atomic cloud location $Z$ (left) and
population $N$ (right) in the unstable zone. In (a), $\Delta _{0}=-0.36$; in
(b), $\Delta _{0}=-0.35$.}
\label{fig:CPSt}
\end{figure}

Fig. \ref{fig:Amplitude} shows that the amplitude of the oscillations
increases with $\Delta _{0}$, so that the maximum value explored by $z$
becomes larger and larger as $\Delta _{0}$ is increased. As a consequence,
the maximum value reached by $z$ also increases, so that finally, the most
distant atoms from the trap center, situated in $z+\Delta z/2$, where $%
\Delta z$ is the size of the cloud, reach the border of the trap, in $z_{0}$%
. In the model, these atoms are considered to be lost, and thus are
subtracted to the total number of atoms in the trap. Therefore, a new
process with a zero characteristic time appears in the model through this
instantaneous decreasing of $n$. This new process leads to an immediate
change of the dynamics frequency. This is illustrated in fig. \ref{fig:freq}%
, where the transition occurs in $\Delta _{0}=-0.361$. For these parameters,
the frequency is divided by more than a factor 3, decreasing from 12 Hz to
3.7 Hz. The new dynamics is illustrated in Fig. \ref{fig:CPSt}. Although the
global shape seems to be similar to the previous $C_{P}$ instabilities, a
drastic difference appears on the dynamics of $N$, in particular concerning
its oscillation amplitude. While the variations of $N$ as a function of time
were small in the $C_{P}$ regime, they appear to be much larger in the
present regime: in fig. \ref{fig:CPSt}a, $N$ varies on half of the total
interval of values that it can take, and when $\Delta _{0}$ is still
increased, this ratio reach 80\%, with an oscillation from 0 to 0.8 (Fig. %
\ref{fig:CPSt}a). In the latter, the cloud empties completely every period,
and then fill up progressively. This explain the large period of the regime,
due to a longer time necessary to fill the cloud. This appears clearly when
fig. \ref{fig:CPt} and fig. \ref{fig:CPSt} are compared: the increasing of
the period does not correspond to a global stretching of the dynamics, as
the large oscillation of $Z$ remains on the same time scale, but rather is a
consequence of the increasing of the interval between two oscillations.

The spectrum confirms that the main frequency has decreased (Fig. \ref%
{fig:CPSfft}). However, this is coupled with the appearance of large
amplitude harmonics, decreasing slowly, so that components with higher
frequency than in the $C_{P}$ regime keep a significant weight.

\begin{figure}[tph]
\centerline{\resizebox{0.45\textwidth}{!}{\includegraphics{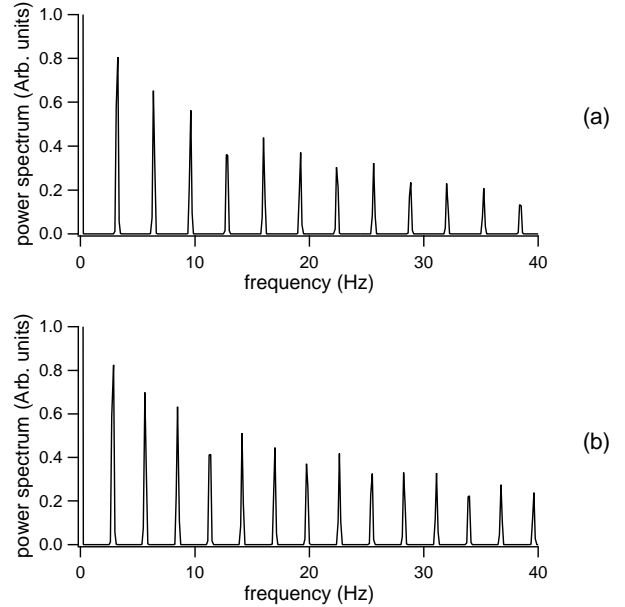}}}
\caption{Power spectra of the atomic cloud location $Z$ in the unstable
zone. Parameters are the same as in Fig.\protect\ref{fig:CPt}.}
\label{fig:CPSfft}
\end{figure}

This behavior has several common points with the $C_{S}$ instabilities
described in the experimental section. In both cases, the regime is the
continuation of the $C_{P}$ instabilities when the resonance is approached,
the amplitude of the $z$ oscillation has a 100\% contrast, that of $n$ are
much larger than in $C_{P}$ regime, the shapes are similar, and new
frequencies of higher value appear. But the regime obtained in the
simulations remains periodic, contrary to that observed experimentally.
However, as discussed in the experimental section, the origin of the
erraticity observed in the experiments could be stochastic, rather than
deterministic. Thus we can hope that the addition on noise in the model will
transform the dynamics to reproduce the experimental results. This influence
of noise on the dynamics is studied in the next section.

\section{The effect of noise}

\label{S7Noise}Noise is known to be able to alter drastically the
deterministic behavior of the MOT: in \cite{bruit2002}, it has been shown
that a stationary behavior may be transformed, under the influence of noise,
in a behavior similar to instabilities. Thus a complete study of the
behaviors predicted by the present model must consider the possible
alterations induced by noise on the dynamics. We present successively in
this section the results obtained on {\it C}$_{{\it P}}$ and {\it C}$_{{\it S%
}}$ instabilities.

\begin{figure}[tph]
\centerline{\resizebox{0.45\textwidth}{!}{\includegraphics{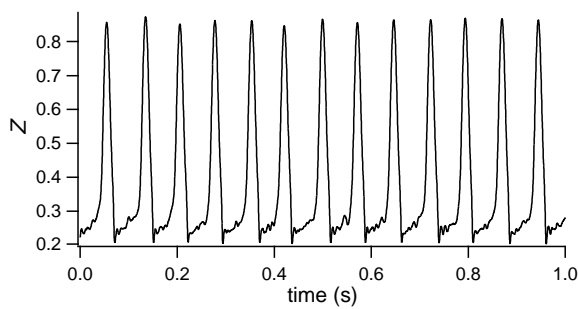}}}
\caption{Time evolution of $Z$ for the same parameters as in Fig.  \protect
\ref{fig:CPt}c, but when noise is added on $I_{1}$. The noise level is 2\%.}
\label{fig:CPtbruit}
\end{figure}

Fig. \ref{fig:CPtbruit} illustrates the effect of noise on {\it C}$_{{\it P}}
$ instabilities, through the example of the regime plotted on Fig. \ref%
{fig:CPt}c with 2\% of noise on $I_{1}$. Although the behavior becomes less
regular, with fluctuating amplitudes and a ruffling of the small secondary
oscillations, the global behavior is unchanged, with a still rather regular
period and the same global shape as without noise. Thus {\it C}$_{{\it P}}$
instabilities appear to be robust against noise. However, if the response to
noise is studied in more details, a slight increase of the noise influence
appears when the {\it C}$_{{\it S}}$ area is approached. This global
behavior allows us to interprete both the {\it C}$_{{\it P}}$ and the {\it C}%
$_{{\it 1}}$ instabilities, which appear in fact to be the same dynamics,
affected differently by noise: between $\Delta _{2}$ and $\Delta _{3}$, i.e.
far from the {\it C}$_{{\it S}}$ area, the {\it C}$_{{\it P}}$ instabilities
are very robust against noise, and the presence of technical noise in the
experiment does not alter neither their shape nor their periodicity. As the 
{\it C}$_{{\it S}}$ area is approached, the sensitivity of the {\it C}$_{%
{\it P}}$ behavior to noise slightly increases, and, although the main
characteristics of the {\it C}$_{{\it P}}$ instabilities remain unchanged,
the shape and periodicity of the regime are affected enough to give the
feeling of a new regime, namely {\it C}$_{{\it 1}}$ instabilities. Thus {\it %
C}$_{{\it 1}}$ instabilities appear, as already suspected in the
experimental section, as a {\it C}$_{{\it P}}$ regime perturbed by noise

\begin{figure}[tph]
\centerline{\resizebox{0.45\textwidth}{!}{\includegraphics{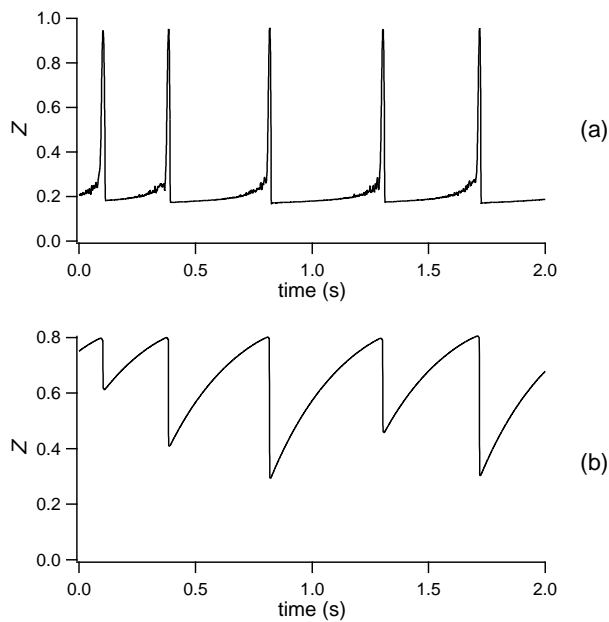}}}
\caption{Time evolution of the (a) atomic cloud location and (b) population,
for the same parameters as in Fig. \protect\ref{fig:CPSt}, but when noise is
added on $I_{1}$. The noise level is 2\%.}
\label{fig:CPStbruit}
\end{figure}

Fig. \ref{fig:CPStbruit} shows how noise alters the dynamics of {\it C}$_{%
{\it S}}$ instabilities. Although the amount of noise is the same as for 
{\it C}$_{{\it P}}$ instabilities, it is clear that here, the dynamics is
deeply transformed. Concerning the $Z$ dynamics, the shape of the
oscillations remains almost unchanged, but the periodicity is drastically
altered: indeed, the return time of the pulses varies randomly on a range
larger than 100\%. The explanation of these large fluctuations on the return
time of the $Z$ pulses comes from the $N$ dynamics: here, the fluctuations
appear on the amplitude of the variable. As the return time is connected to
the reconstruction time of the cloud, it is logical that fluctuations in the
initial population lead to fluctuations in the return time of $Z$. The
strength of the effect is due, as for the stochastic dynamics described in 
\cite{bruit2002}, to an effect of amplification of noise, but through a
different mechanism than that described in \cite{bruit2002}. Indeed, in the
vicinity of $z_{0}$, noise modifies the state of the cloud just before the
brutal decreasing of the population: it will be able to slow down the
crossing of $z_{0}$, or on the contrary to quicken it. The consequence on
the number of atoms lost in the process is immediate, leading to the large
fluctuations of $N$ that can be seen in Fig. \ref{fig:CPStbruit}. If the
resulting dynamics is compared with the experimental one illustrated in Fig. %
\ref{fig:cs}, the similarities between both dynamics appear clearly: the
common points discussed in the previous section remain, and the
discrepancies disappear. In particular, the fluctuations in the return time
of the $z$ pulses, together with those in the amplitude of $n$, are now
present: it is clear that we reproduce here the {\it C}$_{{\it S}}$
instabilities.

This concludes the present study on the deterministic instabilities of the
MOT, as we have been able to retrieve with our model all the behaviors
observed experimentally. In particular, the dynamics that appeared as
erratic is shown to be a deterministic periodic behavior perturbed by noise.
This allows us to do the link with the studies reported in \cite{bruit2002},
as we show in this section that noise plays once again a key role in the
dynamics of the MOT, although here the basis of the dynamics is
deterministic.

\section{Conclusion}

The behavior of the cloud of cold atoms produced by a magneto-optical trap
exhibits a rich variety of dynamics, that are well reproduce by a simple 1D
model described by a set of three autonomous ordinary differential
equations. In \cite{bruit2002} were described a set of noise-induced
instabilities, linked to the topology of the stationary solutions. Here we
show that experimentally, three different regimes of deterministic
instabilities may also be observed, depending on the parameters of the MOT.
Some of these regimes appear to be purely deterministic ({\it C}$_{{\it P}}$
instabilities), some appear to be a mixture of deterministic instabilities
and effects of noise ({\it C}$_{{\it 1}}$ and {\it C}$_{{\it S}}$
instabilities). Theoretically, the same model as in \cite{bruit2002} allows
us to show that the existence of these deterministic instabilities are a
direct consequence of the topological properties that induce for other
parameters the stochastic instabilities studied in \cite{bruit2002}. This
model shows that {\it C}$_{{\it 1}}$ instabilities are just {\it C}$_{{\it P}%
}$ instabilities perturbed by noise, while {\it C}$_{{\it S}}$ instabilities
are a nother deterministic regime appearing when the border of the trap
beams are reached. This last regime is particularly sensitive to noise, and
the resulting behavior appears as a deterministic instability highly
perturbed by noise. Thus the present study confirms that noise plays a
crucial role in the dynamics of the atomic cloud.

The present results have been obtained with a very simple model. Although
the agreement with the experiments is surprisingly good, it is difficult to
make qunatitative comparisons between such a 1D model and a 3D experiment.
Therefore the next theoretical step would be to develop a 3D-model, where
some of the parameters of the present model, as e.g. the cloud volume,
become a function of the dynamics variables.

An interesting perspective of the present results is to study the
possibility to take advantage of the existence of deterministic
instabilities in the MOT. In particular, if a set of parameters could be
found to widen enough the chaotic area, the techniques of control of chaos
could be apply to reach various states that are not accessible otherwise, as
e.g. denser or colder states. But even periodic behaviors can give new
interesting informations about the MOT physics. Indeed, a complex dynamics
covers a larger part of its phase space, and in return makes possible the
determination of parameter values masked in stationary behaviors. More
generally, a complex behavior enables the access to more information about
its system, and appears usually as a good starting point for a better
understanding of it.

\section{Acknowledgments}

The author thanks M. Fauquembergue for her participation in the elaboration
of the model. The Laboratoire de Physique des Lasers, Atomes et Mol\'{e}%
cules is \textquotedblleft Unit\'{e} Mixte de Recherche de l'Universit\'{e}
de Lille 1 et du CNRS\textquotedblright\ (UMR 8523). The Centre d'Etudes et
de Recherches Lasers et Applications (CERLA) is supported by the Minist\`{e}%
re charg\'{e} de la Recherche, the R\'{e}gion Nord-Pas de Calais and the
Fonds Europ\'{e}en de D\'{e}veloppement Economique des R\'{e}gions.


%
%

%

\end{document}